\documentclass[a4paper,12pt,oneside]{article}
\usepackage[cp1251]{inputenc}
\usepackage[english]{babel}
\usepackage{amsmath}
\usepackage{amsfonts}
\usepackage{amssymb}
\usepackage{graphicx}
\usepackage{wasysym}
\usepackage{bm}

\begin{document}
	\begin{center}
		\textbf{Time of falling of a quantum particle into an inverse square potential}\\\bigskip
		
		V.M.~Vasyuta$^1$, V.M.~Tkachuk$^2$\\
		\small\textit{
		E-mail: $^1$waswasiuta@gmail.com, $^2$voltkachuk@gmail.com\\
		Department for Theoretical Physics, Ivan Franko National University of Lviv,\\
			12, Drahomanov St., Lviv, UA-79005, Ukraine}
	\end{center}
	
\begin{quote}\small
	Evolution of a particle in an inverse square potential is studied. We derive an equation of motion for $\left<r^2\right>$ and solve it exactly. It gives us a possibility to identify the conditions under which a falling of a quantum particle into an attractive centre is possible. We get the time of falling of a particle from an initial state into the centre. An example of a quasi-stationary state which evolves with $\left<r^2\right>$ being constant in time is given. We demonstrate the existence of quantum limit of falling, namely, a particle does not fall into the attractive centre, when coupling constant is smaller then some critical value. Our results are compared with experimental measurements of neutral atoms falling in the electric field of a charged wire. Moreover, we propose modifications of the experiment, which allow to observe quantum limit of falling.\bigskip
	
	PACS number(s): 03.65.-w, 03.65.Ca, 03.65.Ge, 37.90.+j, 37.10.Gh.
\end{quote}

\section{Introduction}

The attractive inverse square potential is studied in literature from different points of view. Such interest is motivated by peculiarities of this potential. There are no stationary energy levels for the attractive $1/r^2$ potential, an energy of a particle in this potential is not bounded from below \cite{short,1st,lan}. Formally, one can obtain solutions of the Schr\"{o}dinger equation
\begin{eqnarray}
\label{ham}
	\left(\frac{p^2}{2m}-\frac{\gamma}{r^2}\right)\psi_E=E\psi_E
	\end{eqnarray}
as Bessel functions $Z_\nu(kr)$, where $\nu=\sqrt{1/4+l_z^2/\hbar^2-2m\gamma/\hbar^2}$, $k=\sqrt{2mE/\hbar}$ \cite{short}.
The corresponding functions $\psi_E$ are integrable at the infinity, but in the vicinity of the attractive centre they have asymptotic $\psi_E\sim Ar^{s_+}+Br^{s_-}$, where $s_\pm=-1/2\pm\sqrt{1/4-2m\gamma/\hbar^2}$, and are nonanalytical for a strong coupling regime $\gamma>\hbar^2/8m$. The nonanalytical behaviour of $\psi_E$ can be associated to the classical falling into the centre \cite{lan}. In addition, Hamiltonian (\ref{ham}) is not hermitian \cite{1st}.

Despite the fact that $-1/r^2$ potential is quite exotic, it appears in such physical problems as the Efimov effect \cite{efimov}, a neutral atom in the electric field of a thin charged wire \cite{atwi,dus}, an atom with magnetic moment in the magnetic field of a long solenoid \cite{tk}, the matter near horizon of a black hole \cite{bh}, an electron near a dipolar molecule \cite{elmo}.

Bound states can be restored in an inverse square potential by building self-adjoint extensions of (\ref{ham}) \cite{ext},\cite{rext}. Also bound states for an inverse square potential may appear within the renormalization scheme \cite{renor},\cite{rext}. It is interesting that an inverse square potential is regularized in a natural way in a space with a minimal length \cite{Bou} and noncommutative space \cite{non}. In all these methods an additional parameter with the dimension of length appears, which is not a parameter of $1/r^2$ potential.

The solution of the classical problem is known for a long time, from the early 18th century \cite{cotes}. Trajectory of a particle moving in an inverse square potential is one of the three possible Cotes spirals. A particle can either fall into the attractive centre or escape from it. The situation depends on the value of a coupling constant, z-component of the angular momentum, and initial position.

The aim of this paper is to study the evolution of a quantum particle in an inverse square potential. On the contrary to the classical case, evolution of a quantum particle in an inverse square potential has not been discussed in literature. It is an interesting question how does a wave function evolve under Hamiltonian (\ref{ham}), how does a falling into the centre look for a quantum particle.

Our paper is organized as follows. In Section 2, we consider the evolution of $\left<r^2\right>$ in an inverse square potential. Our theoretical results are compared with experimental measurements of falling of Lithium atoms into the charged wire \cite{dus} in Section 3. In Section 4, we come up with the description of modifications of this experiment, which allows to observe the quantum falling limit experimentally. We summarize our results in the concluding section.

\section{A time of falling}

Calculating an evolution of a wave function under a Hamiltonian with discrete spectrum can be easily done by expanding the wave functions in eigenstates of the Hamiltonian. Doing so, one can act trivially on this expansion by the evolution operator and obtain how the wave function changes over time. But in the case of an inverse square potential, where no stationary levels exist, this scheme cannot be performed.

Another way of calculating the evolution is to expand the evolution operator into the Taylor series by time. For small times this series might converge. But for inverse square potential in this scheme additional problem appears to be related to the singularity of the Hamiltonian. To obtain the evolution of the wave function, one should analyse all terms in Taylor expansion. If the wave function behaves as $r^s$ near the origin, then the $n$-th order term of expansion behaves as $r^s/r^{2n}$ and is singular for $n>s/2$.

The evolution of operator averages can be obtained from Heisenberg equations. Let us consider the evolution of  $\left<r^2\right>$. This average reflects a spatial distribution of the wave function, namely a smaller $\left<r^2\right>$ corresponds to a more compact localization of the wave function, and vice versa. A particle collapses into a point when $\left<r^2\right>=0$. The first derivative of $\left<r^2\right>$ with respect to time reads
\begin{eqnarray}
	\frac{d}{dt}\left<r^2\right>=-\frac{i}{\hbar}\left<\left[r^2,\frac{\bm{p}^2}{2m}+V\right]\right>=\frac{\left<\bm{r}\bm{p}+\bm{p}\bm{r}\right>}{m}.
\end{eqnarray}
The second derivative gives
\begin{eqnarray}
	\frac{d^2}{dt^2}\left<r^2\right>=\frac{4}{m}\left<\frac{\bm{p}^2}{2m}-\frac{1}{2}\bm{r\nabla}V\right>.
\end{eqnarray}

If $V$ is a power function in $r$, one obtain a chain of equations, which can be closed only for three potentials, namely $V=0,\;V=kr^2,\text{ and }V=-\gamma/r^2$. In the case of $V=-\gamma/r^2$ we find
\begin{eqnarray}
\label{eveq}
		\frac{d^2}{dt^2}\left<r^2\right>=\frac{4}{m}\left<H\right>.
\end{eqnarray}

Since the energy is a constant of motion, the solution of (\ref{eveq}) can be immediately found in the following form
\begin{eqnarray}
\label{evor}
	\left<r^2\right>=\left<r^2\right>_0+\frac{\left<\bm{rp}+\bm{pr}\right>_0}{m}t+\frac{2\left<H\right>}{m}t^2,
\end{eqnarray}
where $\left<...\right>_0$ denotes averaging at the initial moment of time $t=0$.

Equation of evolution for $\left<r^2\right>$ (\ref{evor}) coincides with the corresponding classical equation where operator averages are replaced by their classical values. Classical equation can be rewritten in the following manner
\begin{eqnarray}
	\label{clevor}
r^2=r^2_0+2r_0\dot{r}_0t+\frac{2E}{m}t^2.
\end{eqnarray}

The fate of a particle depends on the values of $\left<r^2\right>_0$, $\left<\bm{rp}+\bm{pr}\right>_0$, and $\left<H\right>$. In the case of $\left<H\right><0$ a particle obviously falls into the centre. For $\left<H\right>>0$ a particle either escapes from the attractive centre or falls into the centre. The falling is possible if $\left<\bm{rp}+\bm{pr}\right>_0<0$ and $\left<r^2\right>_0\leq\left<\bm{rp}+\bm{pr}\right>_0^2/8m\left|\left<H\right>\right|$. Inequality $\left<\bm{rp}+\bm{pr}\right>_0<0$ corresponds to the classical case of $\dot{r}_0<0$, i.e. radial projection of the initial velocity is directed to the force centre. An analogous situation is in the case of  $\left<H\right>=0$. It is interesting that for $\left<H\right>=0$ and $\left<\bm{rp}+\bm{pr}\right>_0=0$ the particle moves with a constant average $\left<r^2\right>$. The time dependence of $\left<r^2\right>$ is plotted in the Figure 1. Falling time can be calculated from (\ref{evor}) by putting $\left<r^2\right>=0$.
\begin{figure}[h!]
\begin{center}
	\includegraphics[clip,scale=0.47]{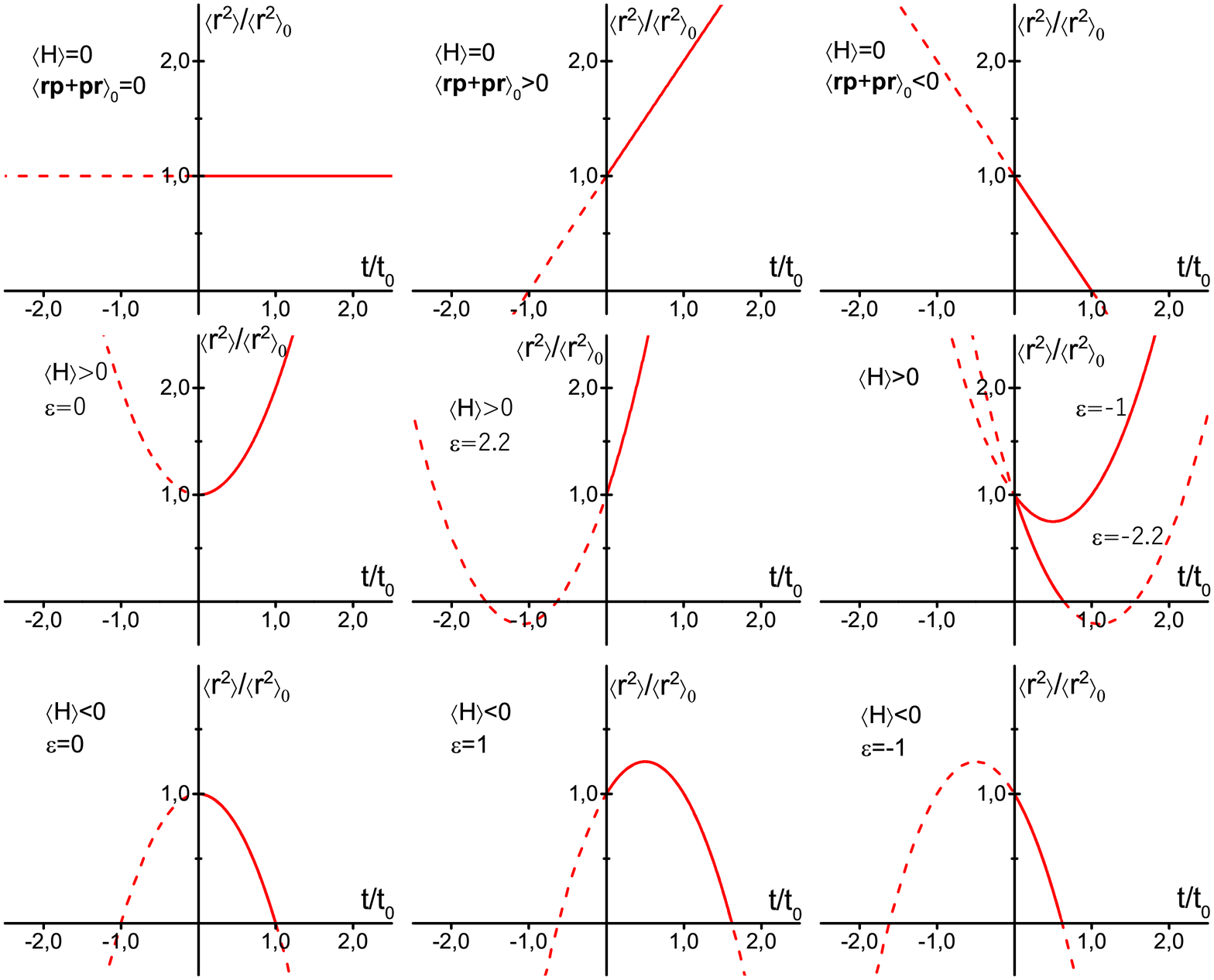}\\
\end{center}
	FIG. 1. Evolution of $\left<r^2\right>$ for different initial conditions. The complete line (dashed and solid) represents the solution of the equation of motion for $\left<r^2\right>$, but a real motion of a particle is pictured by the solid part. The time scale factor $t_0$ is defined in the following way. If $\left<H\right>\neq0$, then $t_0=\sqrt{m\left<r^2\right>_0/2\left|\left<H\right>\right|}$ and $\varepsilon=\left<\bm{rp+pr}\right>_0/\sqrt{2m\left|\left<H\right>\right|\left<r^2\right>_0}$. In the case of $\left<H\right>=0$ we can put $t_0=m\left<r^2\right>_0/\left|\left<\bm{rp+pr}\right>_0\right|$. If both $\left<H\right>$ and $\left<\bm{rp+pr}\right>_0$ equal zero,  particle evolve with a constant $\left<r^2\right>$, and in this case the unit of time can be chosen arbitrarily.
\end{figure}

In the next section we will compare our theoretical results with measurements of neutral atoms motion in the electric field of a charged wire \cite{dus} (from here and through the text we will call it as DUS experiment due to names of authors). In this experiment particles move in two-dimensional potential. So, the 2D case will be considered in detail. Due to the operator identity $\bm{rp}+\bm{pr}=r\frac{\partial}{\partial r}+\frac{\partial}{\partial r}r$, in 2D space when the radial part of the wave function is real we have
\begin{eqnarray}
\label{cond1}
\left<\bm{rp}+\bm{pr}\right>_0=0.
\end{eqnarray}

Taking into account condition (\ref{cond1}), we can obtain time of falling as follows
\begin{eqnarray}
\label{ft}
	t_f=\sqrt{-\frac{m\left<r^2\right>_0}{2\left<H\right>}}.
\end{eqnarray}
If (\ref{cond1}) holds, the particle falls into the centre only if its energy is negative. If the wave function is proportional to $r^s$ near the origin with $s<1/2$, $\left<H\right>$ formally becomes infinite, because corresponding integrand is singular. In this case, the time of falling equals zero. But such states require an infinite energy for creation, therefore they are unphysical. For positive energies $t_f$ is imaginary, and particle cannot fall into the centre. 

It is interesting that a particle with zero energy will stay infinitely long on the same distance from the centre $\left<r^2\right>=\left<r^2\right>_0$; this fact is easily seen from (\ref{evor}). In this case, the time of falling formally is infinite. An example of such a quasi-stationary state can be found in the following form
\begin{eqnarray}
\label{exampl}
\psi_s=\sqrt{\frac{2\beta^{s+1}}{\Gamma(s+1)}}r^s\exp(-\frac{1}{2}\beta r^2),
\end{eqnarray}
where $\Gamma(s)$ is the Euler's gamma function. The energy is equal to zero for $s=2m\gamma/3\hbar^2$. This condition provides the quasi-stationary motion.

Classical expression for a falling time can be obtained from the quantum one (\ref{ft}) by replacing operator averages by their classical values. The principal difference between classical and quantum evolution is in a set of allowed values of classical quantities and corresponding quantum mechanical averages. An interesting quantum effect appears when $\gamma$ is small. The classical particle can fall into the centre for arbitrarily small $\gamma$. If one put $\dot{r}=0\text{ and }L=0$ then $E$ becomes negative, the particle will fall into the centre. In quantum mechanics situation is quite different, there is a critical value $\gamma_c$, so that for $\gamma\leq\gamma_c$ particle will not fall into the centre. This quantum limit exists due to the fact that for small $\gamma$ an energy of a particle cannot be negative. To obtain this critical value $\gamma_c$ let us analyse the average of Hamiltonian in detail. For the sake of simplicity, we take the wave function in the form $\psi=\mathcal{R}(r)e^{il_z\varphi}/\sqrt{2\pi}$, where radial part $\mathcal{R}(r)$ is an arbitrary function, the angular part is an eigenfunction of orbital momentum, which corresponds to the eigenvalue $l_z$. In this case one can rewrite $\left<H\right>$ in the following form
\begin{eqnarray}
\label{avham}
\left<H\right>=\frac{\left<p_r^2\right>}{2m}+\left[\frac{\hbar^2}{2m}\left(l_z^2+\frac{1}{4}\right)-\gamma\right]\left<\frac{1}{r^2}\right>,
\end{eqnarray}
where $p_r=\frac{1}{\sqrt{r}}\partial_r\sqrt{r}$. Using the condition $\left<H\right><0$, we get
\begin{eqnarray}
\label{gc}
	\gamma>\gamma_c=\frac{\left<p_r^2\right>}{2m\left<1/r^2\right>}+\frac{\hbar^2}{2m}\left(l_z^2+\frac{1}{4}\right).
\end{eqnarray}

From (\ref{gc}) it follows that  $\min\gamma_c=\Gamma_c=\frac{\hbar^2}{8m}$. It is remarkable that $\Gamma_c$ coincides with the limit of a strong coupling regime. For a given wave function the critical value $\gamma_c$ can be calculated from (\ref{gc}). For example, for the wave function $\psi_s$ (\ref{exampl}) (integrand is nonsingular when $s\geq1/2$) we obtain
\begin{eqnarray}
\label{dusft}
\left<H\right>=\frac{s+1}{\left<r^2\right>_0}\left(\frac{\hbar^2}{2m}-\frac{\gamma}{s}\right),
\end{eqnarray}
and for the critical value $\gamma_c$ we get
\begin{eqnarray}
\label{dusft2}
\gamma_c=\frac{s\hbar^2}{2m}=4s\Gamma_c.
\end{eqnarray}

\section{Neutral atoms in the field of a thin charged wire}

Falling into an inverse square potential was experimentally realized in DUS experiments with ultracold Lithium atoms in the field of a thin charged wire \cite{dus}. In this experiment neutral Lithium atoms with polarizability $\alpha_{Li}$ were loaded in a cylindrical chamber with length of $l=0.1\text{ m}$. The thin wire with a radius of $r_1=0.7\text{ }\mu\text{m}$ was placed coaxially with the chamber. Radius of the external cylinder is not pointed out in papers, but it can be obtained using the relation between the charge on the wire and the voltage on it (external cylinder is grounded): a charge of $q=640\;\frac{\text{pC}}{\text{m}}$ corresponds to the voltage of $U=100\text{ V}$. Using the expression for the capacitance of a cylindrical capacitor, we obtain the chamber radius as $r_2=4.2\text{ mm}$. In this experiment the authors achieved the pressure of $p=6\cdot10^{-10}$ Torr. This pressure is sufficiently small, thereby we can assume that Lithium atoms do not collide with each other during the motion.

Neutral atoms are polarized by the electric field of a wire and obtain an electric dipole moment $\varepsilon_0\alpha E$. The Hamiltonian of the induced electrical dipole in an electric field $E$ reads
\begin{eqnarray}
	H=\frac{p^2}{2m}-\frac{\varepsilon_0\alpha E^2}{2}=\frac{p^2}{2m}-\frac{\gamma}{r^2},
\end{eqnarray}
where $\gamma=\alpha q^2/8\pi^2\varepsilon_0$.

Now we want to estimate the time of falling of atoms in this chamber using expression (\ref{ft}). Let us assume that the wave function of the Lithium atoms is equal to (\ref{exampl}) with $s=1/2$. Putting $\left<r^2\right>=r_2^2$ and using (\ref{ft}) and (\ref{dusft}), we obtain the time of falling of about $t_f\sim7\text{ s}$. This result is in correspondence with the estimation made in \cite{dus}, where the lifetime of atoms of about $10\text{ s}$ was measured.

An exact wave function of Lithium atoms is not known. So, to estimate a credible value of critical charge $q_c$ needed for a falling of quantum particle we have to use expression for $\Gamma_c$. Finally, we obtain
\begin{eqnarray}
\label{qc}
	q_c=\sqrt{\frac{\hbar^2\pi^2\varepsilon_0}{m\alpha}}.
\end{eqnarray}
In the classical limit ($\hbar\rightarrow0$) the critical charge vanishes $q_c=0$, as it should be.

The critical charge of the wire in DUS experiment, for which Lithium atoms ($\alpha_{Li}=24.3\text{ \AA}^3$) do not fall into the centre, is about a $q=1.8\;\frac{\text{pC}}{\text{m}}$. This charge corresponds to a voltage of $U_c=0.29\text{ V}$. Because of a noise produced by optical trap in the chamber such value of voltage cannot be measured in this experiment.

\section{Proposed experiment for observing quantum limit}

The aim of this section is to give a proposal how to increase the critical value $U_c$ and make it observable. The first proposal is to use lighter atoms with smaller value of polarizability, as it follows from expression for $q_c$ (\ref{qc}). For Hydrogen atoms ($\alpha_H=0.667\text{ \AA}^3$) the critical charge and voltage for the chamber in the DUS experiment are $q_c=30\;\frac{\text{pC}}{\text{m}}$, $U_c=4.6\text{ V}$ respectively. So, for Hydrogen atoms values of quantum limits are 16 times bigger, than for Lithium atoms. Almost the same values are obtained for $^3$He atoms: $q=31\;\frac{\text{pC}}{\text{m}}$, $U_c=4.8\text{ V}$.

Another idea is to change a geometry of the chamber. This proposal is motivated by the relation between the charge and the voltage of capacitor
\begin{eqnarray}
	U=\frac{q}{C}=\frac{q}{2\pi\varepsilon_0}\ln\frac{r_2}{r_1},
\end{eqnarray}
where $C$ is the capacitance. So, using a capacitor with bigger $r_2$ and smaller $r_1$ allows us to work with bigger voltage. For another capacitor with $r_2'=\lambda_2r_2,\;r_1'=\lambda_1r_1$ the value of the critical voltage reads
\begin{eqnarray}
	U_c'=U_c\frac{\ln\frac{\lambda_2r_2}{\lambda_1r_1}}{\ln\frac{r_2}{r_1}}=U_c\left(1+\frac{\ln \frac{\lambda_2}{\lambda_1}}{\ln\frac{r_2}{r_1}}\right).
\end{eqnarray}

For DUS experiment we have $\ln\frac{r_2}{r_1}\approx8.67$, so the critical voltage is $U_c'\approx1.26U_c$ for $\lambda_2/\lambda_1=10$ ($r_2=4.2\text{ cm}$, $r_2=0.7\;\mu\text{m}$) and $U_c'\approx1.53U_c$ for $\lambda_2/\lambda_1=100$ ($r_2=42\text{ cm}$, $r_2=0.7\;\mu\text{m}$). Increasing the size of chamber by 10 times ensures the critical voltage of about $6.1\text{ V}$ for $^3$He atoms and $5.8\text{ V}$ for H atoms. It is almost 21 times bigger voltage than the critical voltage in the original DUS experiment. We hope that such improvements give a possibility to observe the quantum limit of falling experimentally.

\section{Conclusions}

We have shown that an average $\left<r^2\right>$ evolves as a quadratic polynomial in time in an inverse square potential. This evolution is fully defined by averages of $\left<r^2\right>_0$, $\left<\bm{rp}+\bm{pr}\right>_0$, and $\left<H\right>$ in the initial state and is qualitatively different for different initial conditions (Fig. 1). We have found necessary conditions needed for a falling of a quantum particle into the attractive centre. The case of $\left<\bm{rp}+\bm{pr}\right>_0=0$ in two-dimensional space has been considered in detail. For such states the time of falling is given by expression (\ref{ft}). We have shown that there are quasi-stationary states, which evolve with constant $\left<r^2\right>$. Also an example of such a state has been given (\ref{exampl}). By analysing the average $\left<H\right>$, we have established a quantum limit of falling, namely a particle cannot fall into the attractive centre if the coupling constant $\gamma$ is smaller than some critical value $\gamma_c$. The critical value $\gamma_c$ is defined by the expression (\ref{gc}).

We have compared the obtained results with the experimental measurements of the motion of neutral Lithium atoms in the electric field of a charged wire \cite{dus}. Calculated theoretically time of falling coincides with the experimentally measured value. Also, we have calculated the critical charge of the wire and the voltage between the wire and the wall of a chamber, which allows quantum particle to fall into the centre. The critical voltage is about 0.3 V.

Unfortunately, critical parameters of quantum falling of Lithium atoms in DUS experiment are so small that they could not be measured there. We have given some proposals how to improve the experiment to observe the quantum falling limit. Namely, we propose to use lighter atoms with smaller polarizability (Hydrogen or Helium atoms) and to change the size of a chamber. This improvements make a voltage about $6.1\text{ V}$, that is almost 21 times bigger than analogous values in DUS experiment. We hope that these improvements allow to observe the quantum limit of falling in the experiment.


\begin{thebibliography}{9}
\bibitem{short} G. H. Shortley, Phys. Rev. \textbf{38}, 120 (1931).
\bibitem{1st} K. M. Case, Phys. Rev. \textbf{80}, 797 (1950); 
E. A. Guggenheim, Proc. Phys. Soc. \textbf{89}, 491 (1966).
\bibitem{lan} L. D. Landau and E. M. Lifshitz, Quantum Mechanics: Nonrelativistic Theory (Fizmatlit, Moscow, 2004).
\bibitem{efimov} V. N. Efimov, Sov. J. Nucl. Phys. \textbf{12}, 589 (1971).
\bibitem{atwi} L. V. Hau, M. M. Burns, and J. A. Golovchenko, Phys. Rev. A \textbf{45}, 6468 (1992);
J. Schmiedmayer, Appl. Phys. B \textbf{60}, 169 (1995);
J. Denschlag and J. Schmiedmayer, Europhys. Lett. \textbf{38}, 405 (1997).
\bibitem{dus} J. Denschlag, G. Umshaus, and J. Schmiedmayer, Phys. Rev. Lett. \textbf{81}, 737, (1998).
\bibitem{tk} V. M. Tkachuk, Phys. Rev. A \textbf{60}, 4715, (1999).
\bibitem{bh} T. R. Govindarajan, V. Suneeta, and S. Vaidya, Nucl. Phys. B
\textbf{583}, 291 (2000);
D. Birmingham, K. S. Gupta, and S. Sen, Phys. Lett. B \textbf{505}, 191 (2001);
K. S. Gupta and S. Sen, Phys. Lett. B \textbf{526}, 121 (2002);
S. K. Chakrabarti, K. S. Gupta, and S. Sen, Int. J. Mod. Phys. A \textbf{23}, 2547 (2008);
H. E. Camblong and C. R Ord\'{o}\~{n}ez, Class. Quantum Grav. \textbf{30}, 175007 (2013).
\bibitem{elmo}M. Bawin, Phys. Rev. A \textbf{70}, 022505 (2004);
M. Bawin, S. A. Coon, and B. R. Holstein, Int. J. Mod. Phys. A \textbf{22}, 4901 (2007); A. D. Alhaidari, J. Phys. A: Math.
Theor. \textbf{40}, 14843 (2007);
P. R. Giri, K. S. Gupta, S. Meljanac, and A. Samsarov, Phys. Lett. A, \textbf{372}, 2967, (2008).
\bibitem{ext}H. Narnhofer, Acta Phys. Austriaca \textbf{40}, 306 (1974);
M. Bawin and S. A. Coon, Phys. Rev. A \textbf{67}, 042712 (2003).
\bibitem{renor}K. S. Gupta and S. G. Rajeev, Phys. Rev. D \textbf{48}, 5940 (1993);
H. E. Camblong, L. N. Epele, H. Fanchiotti, and C. A. Garc\'{i}a
Canal, Phys. Rev. Lett. \textbf{85}, 1590 (2000);
S. R. Beane, P. F. Bedaque, L. Childress, A. Kryjevski,
J. McGuire, and U. van Kolck, Phys. Rev. A \textbf{64}, 042103 (2001);
S. A. Coon and B. R.
Holstein, Am. J. Phys. \textbf{70}, 513 (2002); H.-W. Hammer and B.
G. Swingle, Ann. Phys. \textbf{321}, 306 (2006);
A. D. Alhaidari, Found. Phys. \textbf{44}, 1049 (2014), and references therein.
\bibitem{rext} D. Bouazis and M. Bawin, Phys. Rev. A \textbf{89}, 022113 (2014), and references therein.
\bibitem{Bou} D. Bouaziz and M. Bawin, Phys. Rev. A \textbf{76}, 032112 (2007);
D. Bouaziz and M. Bawin, Phys. Rev. A \textbf{78}, 032110 (2008).
\bibitem{non} P. R. Giri, Int. J. Mod. Phys. A, \textbf{24}, 2655 (2009).
\bibitem{cotes}R. Cotes, Harmonia Mensurarum (Cambridge, 1722).

\end{thebibliography}
\end{document}